\documentclass[11pt]{article}
\usepackage{times}
\usepackage{graphicx}

\begin{document}


\title{300x Faster Matlab using MatlabMPI
\thanks{
This work is sponsored by the High Performance Computing Modernization
Office, under Air Force Contract
F19628-00-C-0002.  Opinions, interpretations, conclusions and
recommendations are those of the author and are not necessarily endorsed
by the Department of Defense.
}}

\author{Jeremy Kepner  (kepner@ll.mit.edu) \\
MIT Lincoln Laboratory, Lexington, MA  02420 \\
Stan Ahalt (sca@ee.eng.ohio-state.edu) \\
Department of Electrical Engineering \\
The Ohio State University, Columbus, OH 43210 \\
}

\date{May 3, 2002}
\maketitle

\begin{abstract}

  The true costs of high performance computing are currently dominated
by software.  Addressing these costs requires shifting to high
productivity languages such as Matlab.  MatlabMPI is a Matlab
implementation of the Message Passing Interface (MPI) standard and
allows any Matlab program to exploit multiple processors. MatlabMPI
currently implements the basic six functions that are the core of the
MPI point-to-point communications standard. The key technical
innovation of MatlabMPI is that it implements the widely used MPI
``look and feel'' on top of standard Matlab file I/O, resulting in an
extremely compact ($\sim$250 lines of code) and ``pure''
implementation which runs anywhere Matlab runs, and on any heterogeneous
combination of computers.  The performance has been tested on both
shared and distributed memory parallel computers (e.g. Sun, SGI, HP,
IBM and Linux).  MatlabMPI can match the bandwidth of C based MPI at
large message sizes. A test image filtering application using
MatlabMPI achieved a speedup of $\sim$300 using 304 CPUs and
$\sim$15\% of the theoretical peak (450 Gigaflops) on an IBM SP2 at
the Maui High Performance Computing Center.  In addition, this entire
parallel benchmark application was implemented in 70
software-lines-of-code (SLOC) yielding 0.85 Gigaflop/SLOC or 4.4
CPUs/SLOC, which are the highest values of these software price
performance metrics ever achieved for any application.  The MatlabMPI
software will be made available for download.

\end{abstract}

\section{Introduction}

  Matlab \cite{Matlab} is the dominant programming language for
implementing numerical computations and is widely used for algorithm
development, simulation, data reduction, testing and system
evaluation.  The popularity of Matlab is driven by the high
productivity that is achieved by users because one line of Matlab code
can typically replace ten lines of C or Fortran code.  Many Matlab
programs can benefit from faster execution on a parallel computer, but
achieving this goal has been a significant challenge. There have been
many previous attempts to provide an efficient mechanism for running
Matlab programs on parallel computers
\cite{MATABP,Morrow98,ParAl,RTExpress,Tseng99,MultiMATLAB,
      ParaMat,Fabozzi99,Matpar,MPITB,Quinn,CMTM}.
These efforts of have
faced numerous challenges and none have received widespread
acceptance.

  In the world of parallel computing the Message Passing Interface
(MPI) \cite{MPI} is the de facto standard for implementing programs on
multiple processors. MPI defines C and Fortran language functions for
doing point-to-point communication in a parallel program.  MPI has
proven to be an effective model for implementing parallel programs and
is used by many of the worlds' most demanding applications (weather
modeling, weapons simulation, aircraft design, and signal processing
simulation).

  MatlabMPI consists of a set of Matlab scripts that implements a subset
of MPI and allows any Matlab program to be run on a parallel computer.
The key innovation of MatlabMPI is that it implements the widely used
MPI ``look and feel'' on top of standard Matlab file I/O, resulting in a
``pure'' Matlab implementation that is exceedingly small ($\sim$250 lines
of code). Thus, MatlabMPI will run on any combination of computers that
Matlab supports.

  The next section describes the implementation and functionality provided
by MatlabMPI.  Section three presents results on the bandwidth performance
of the library from several parallel computers.  Section four uses an image
processing application to show the scaling performance and the very
high software price performance that can be achieved using MatlabMPI.
Section five presents our conclusions and plans for future work.

\section{Implementation}

\begin{figure}[tbh]
\centerline{\includegraphics[width=4.5in]{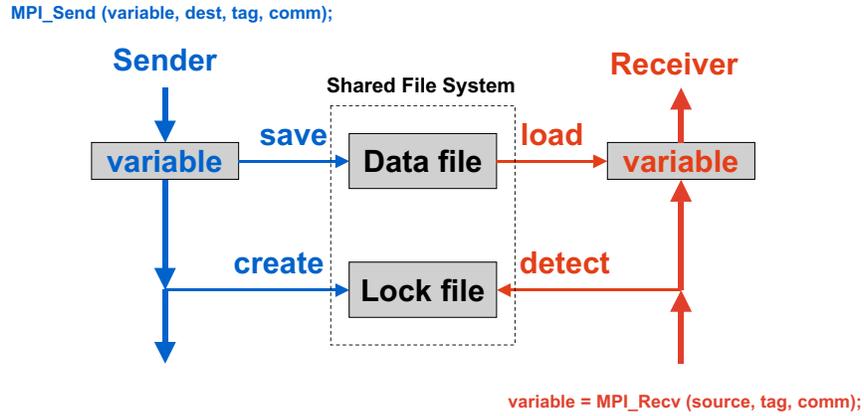}}
\caption{ {\bf File Based Communication.}
Sender saves a variable in a Data file, then creates Lock file.
Receiver detects Lock file, then loads Data file.
}
\label{fig:FileBasedComm}
\end{figure}

  The central innovation of MatlabMPI is the use of Matlab's built
in file I/O capabilities, which allow any Matlab variable (matrices,
arrays, structures, ...) to be written and read by Matlab running on
any machine, and eliminates the need to write complicated buffer
packing and unpacking routines.  The approach used in MatlabMPI is
extremely simple and is illustrated in Figure
~\ref{fig:FileBasedComm}.  The sender saves a Matlab variable
to a data file and when the file is complete the sender touches
a lock file.  The receiver continuously checks for the existence
of the lock file; when it is exists the receiver reads in the data
file and then deletes both the data file and the lock file.  An
example of a basic send and receive Matlab program is shown below
\begin{quote}
\begin{verbatim}
MPI_Init;                           % Initialize MPI.
comm = MPI_COMM_WORLD;              % Create communicator.
comm_size = MPI_Comm_size(comm);    % Get size.
my_rank = MPI_Comm_rank(comm);      % Get rank.
source = 0;                         % Set source.
dest = 1;                           % Set destination.
tag = 1;                            % Set message tag.
if(comm_size == 2)                  % Check size.
  if (my_rank == source)            % If source.
    data = 1:10;                    % Create data.
    MPI_Send(data,dest,tag,comm);   % Send data.
  end
  if (my_rank == dest)              % If destination.
    data=MPI_Recv(source,tag,comm); % Receive data.
  end
end
MPI_Finalize;                       % Finalize Matlab MPI.
exit;                               % Exit Matlab
\end{verbatim}
\end{quote}
The structure of the above program is very similar to MPI programs
written in C or fortran, but with the convenience of a high level
language.  The first part of the program sets up the MPI world; the
second part differentiates the program according to rank and executes
the communication; the third part closes down the MPI world and
exits Matlab.  If the above program were written in a matlab file
called {\tt SendReceive.m}, it would be executed by calling the
following command from the Matlab prompt:
\begin{verbatim}
    eval( MPI_Run('SendReceive',2,machines) );
\end{verbatim}
Where the {\tt machines} argument can be any of the following forms:
\begin{quote}
\begin{description}
   \item[{\tt machines = \{\};}]
     Run on the local host.
   \item[{\tt machines = \{'machine1' 'machine2'\};}]
     Run on a multi processors.
    \item[{\tt machines = \{'machine1:dir1' 'machine2:dir2'\};}]
      Run on a multi processors and communicate using dir1 and dir2,
      which must be visible to both machines.
\end{description}
\end{quote}
The {\tt MPI\_Run} command launches a Matlab script on the specified
machines with output redirected to a file.  If the rank=0 process is
being run on the local machine, {\tt MPI\_Run} returns a string
containing the commands to initialize MatlabMPI, which allows
MatlabMPI to be invoked deep inside of a Matlab program in a manner
similar to fork-join model employed in OpenMP.

  The {\tt SendReceive} example illustrates the basic six MPI
functions (plus {\tt MPI\_Run}) that have been implemented in
MatlabMPI
\begin{quote}
\begin{description}
  \item[{\tt MPI\_Run}]
    Runs a matlab script in parallel.
  \item[{\tt MPI\_Init}]
    Initializes MPI.
  \item[{\tt MPI\_Comm\_size}]
    Gets the number of processors in a communicator.
  \item[{\tt MPI\_Comm\_rank}]
    Gets the rank of current processor within a communicator.
  \item[{\tt MPI\_Send}]
    Sends a message to a processor (non-blocking).
  \item[{\tt MPI\_Recv}]
    Receives message from a processor (blocking).
  \item[{\tt MPI\_Finalize}]
    Finalizes MPI.
\end{description}
\end{quote}
For convenience, three additional MPI functions have also been
implemented along with two utility functions that provide
important MatlabMPI functionality, but are outside
the MPI specification
\begin{quote}
\begin{description}
  \item[{\tt MPI\_Abort}]
    Function to kill all matlab jobs started by MatlabMPI.
  \item[{\tt MPI\_Bcast}]
    Broadcast a message (blocking).
  \item[{\tt MPI\_Probe}]
    Returns a list of all incoming messages.
  \item[{\tt MatMPI\_Save\_messages}]
    MatlabMPI function to prevent messages
    from being deleted (useful for debugging).
  \item[{\tt MatMPI\_Delete\_all}]
    MatlabMPI function to delete all files created by MatlabMPI.
\end{description}
\end{quote}

  MatlabMPI handles errors the same as Matlab, however running
hundreds of copies does bring up some additional issues.  If an error
is encountered and the Matlab script has an ``exit'' statement then
all the Matlab processes will die gracefully.  If a Matlab job is
waiting for a message that will never arrives than it needs to be
killed with the {\tt MPI\_Abort} command.  In this situation,
MatlabMPI can leave files which need to be deleted with the {\tt
MatMPI\_Delete\_all} command.

  On shared memory systems, MatlabMPI only requires a single Matlab
license since any user is allowed to launch many Matlab sessions. On a
distributed memory system, MatlabMPI requires one Matlab license per
machine. Because MatlabMPI uses file I/O for communication, there must
also be a directory that is visible to every machine (this is usually
also required in order to install Matlab).  This directory defaults to
the directory that the program is launched from.

\section{Bandwidth}

  The vast majority of potential Matlab applications are
``embarrassingly'' parallel and require minimal performance out of
MatlabMPI.  These applications exploit coarse grain parallelism and
communicate rarely (if at all). Never-the-less, measuring the
communication performance is useful for determining which applications
are most suitable for MatlabMPI.

\begin{figure}[tbh]
\centerline{\includegraphics[width=4.5in]{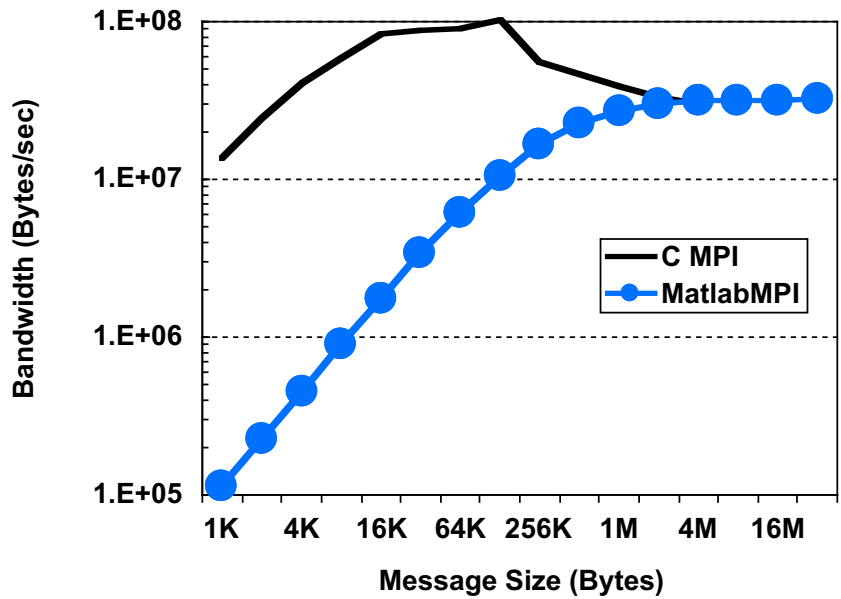}}
\caption{ {\bf MatlabMPI vs. MPI Bandwidth.}
  Communication performance as a function message size on the SGI Origin2000.
MatlabMPI equals C MPI performance at large message sizes.
}
\label{fig:BU_O2000_bandwidth}
\end{figure}

\begin{figure}[tbh]
\centerline{\includegraphics[width=4.5in]{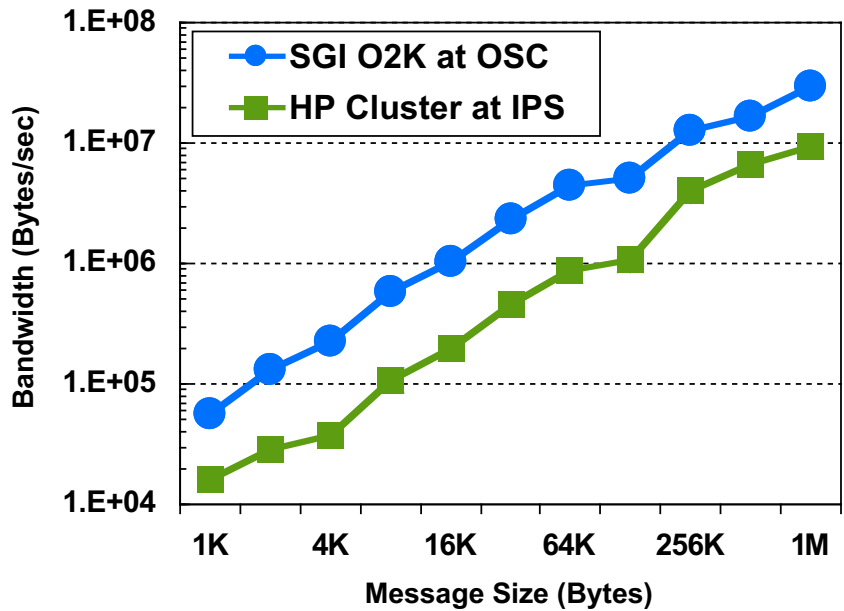}}
\caption{ {\bf Bandwidth Comparison.}
  Bandwidth measured on Ohio St. SGI Origin2000 and a Hewlett
Packard workstation cluster.
}
\label{fig:OSU_HP_O2000_bandwidth}
\end{figure}

  MatlabMPI has been run on several Unix platforms.  It has been
benchmarked and compared to the performance of C MPI on the SGI
Origin2000 housed at Boston University.  These results indicate that
for large messages ($\sim$1 MByte) MatlabMPI is able to match the
performance of C MPI (see Figure~\ref{fig:BU_O2000_bandwidth}).  For
smaller messages, MatlabMPI is dominated by its latency ($\sim$35
milliseconds), which is significantly larger than C MPI.  These
results have been reproduced using a SGI Origin2000 and a Hewlett
Packard workstation cluster (connected with 100 Mbit ethernet) at Ohio
State University (see Figure~\ref{fig:OSU_HP_O2000_bandwidth}).

\begin{figure}[tbh]
\centerline{\includegraphics[width=4.5in]{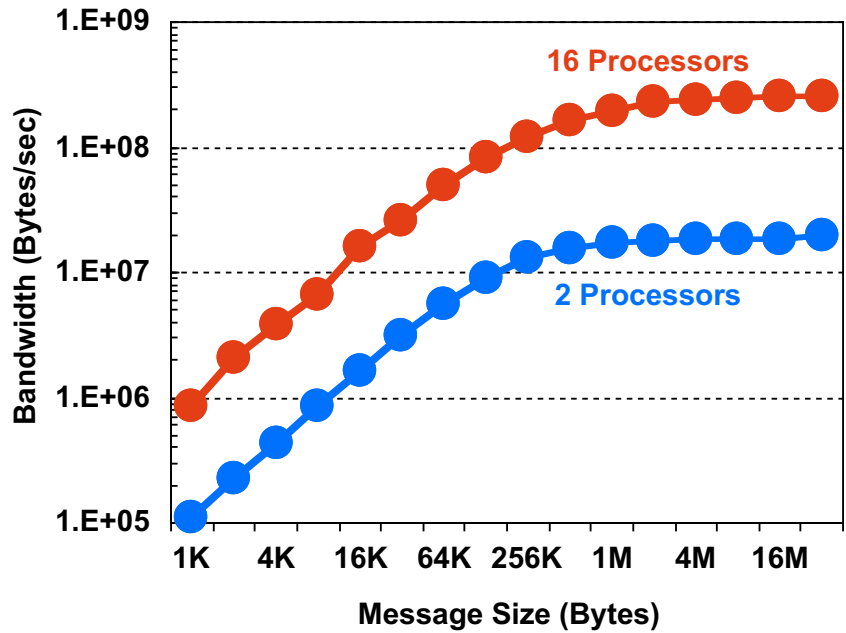}}
\caption{ {\bf Bandwidth on a Linux Cluster.}
Communication performance on Linux cluster with Gigabit ethernet.
MatlabMPI is able to maintain communication performance even when
larger numbers of processor are communicating.
}
\label{fig:Linux_bandwidth}
\end{figure}

  The above bandwidth results were all obtained using two processors
engaging in bi-directional sends and receives.  Such a test does a
good job of testing the individual links on an a multi-processor
system.  To more broadly test the interconnect the send receive
benchmark is run on an eight node (16 cpu) Linux cluster connected
with Gigabit ethernet (see Figure~\ref{fig:Linux_bandwidth}).  These
results are shown for one and 16 cpus.  MatlabMPI is able to maintain
high bandwidth even when multiple processors are communicating by
allowing each processor to have its own receive directory.  By cross
mounting all the disks in a cluster each node only sees the traffic
directed to it, which allows communication contention to be kept to a
minimum.

\section{Scaling}

\begin{figure}[tbh]
\centerline{\includegraphics[width=4.5in]{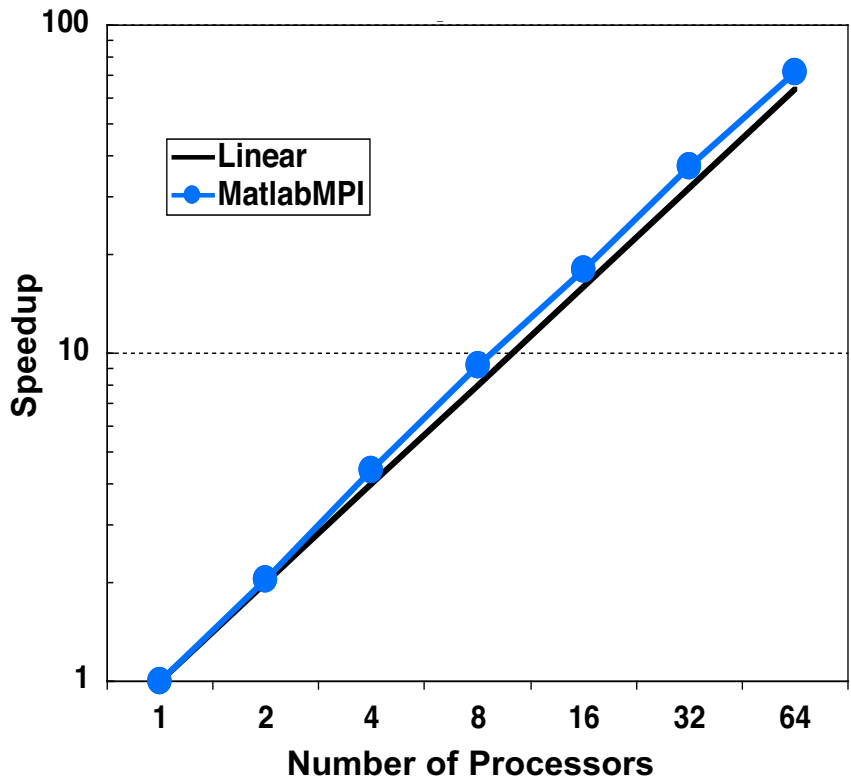}}
\caption{ {\bf Shared Memory Parallel Speedup.}
  Speed increase on the SGI Origin2000 of a parallel image filtering
application as a function of the number of processors. Application shows
``classic'' super-linear performance (due to better cache usage) that
results when a very large memory problem is broken into multiple small
memory problems.
}
\label{fig:BU_O2000_speedup}
\end{figure}

  To further test MatlabMPI a simple image filtering application was
written. This application abstracts the key computations that are used
in many of our DoD sensor processing applications (e.g. wide area
Synthetic Aperture Radar).  The application executed repeated 2D
convolution on a large image (1024 x 128,000 $\sim$2~GBytes).  This
application was run on a large shared memory parallel computer (the SGI
Origin2000 at Boston University) and achieved speedups greater than 64
on 64 processors; showing the classic super-linear speedup (due to
better cache usage) that comes from breaking a very large memory
problem into many smaller problems (see
Figure~\ref{fig:BU_O2000_speedup}).

\begin{figure}[tbh]
\centerline{\includegraphics[width=4.5in]{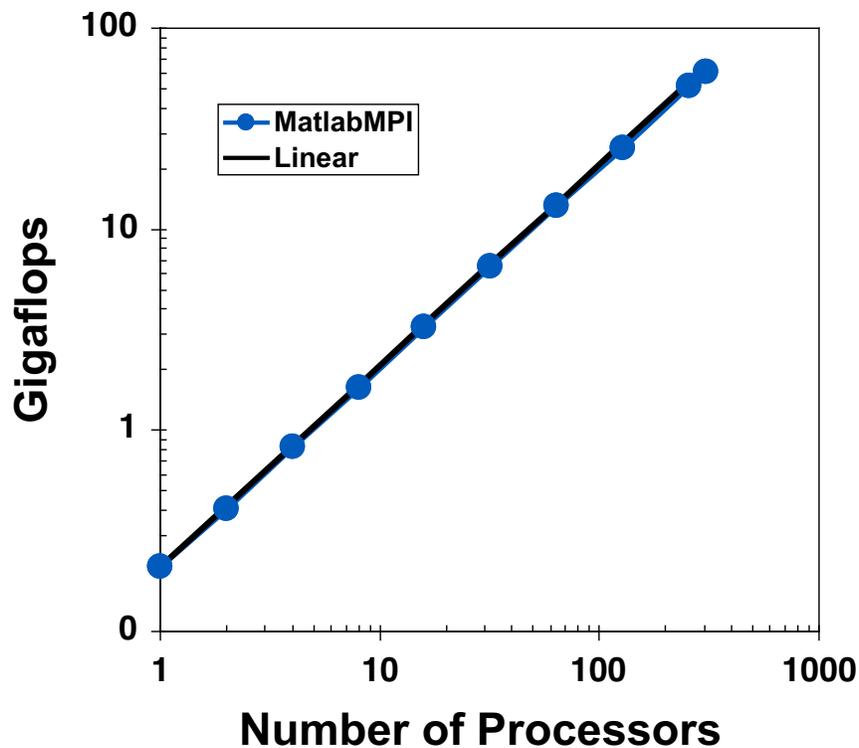}}
\caption{ {\bf Shared/Distributed Parallel Speedup.}
Measure performance on the IBM SP2 of a parallel image filtering
application.  Application achieves a speedup of $\sim$300 on 304
processors and $\sim$15\% of the theoretical peak (450 Gigaflops).
}
\label{fig:MHPCC_IBMSP2_speedup}
\end{figure}

  To further test the scalability, the image processing application
was run with a constant load per processor (1024 x 1024 image per
processor) on a large shared/distributed memory system (the IBM SP2 at the
Maui High Performance Computing Center).  In this test, the
application achieved a speedup of $\sim$300 on 304 CPUs as well
achieving $\sim$15\% of the theoretical peak (450 Gigaflops) of the
system (see Figure~\ref{fig:MHPCC_IBMSP2_speedup}).

\begin{figure}[tbh]
\centerline{\includegraphics[width=4.5in]{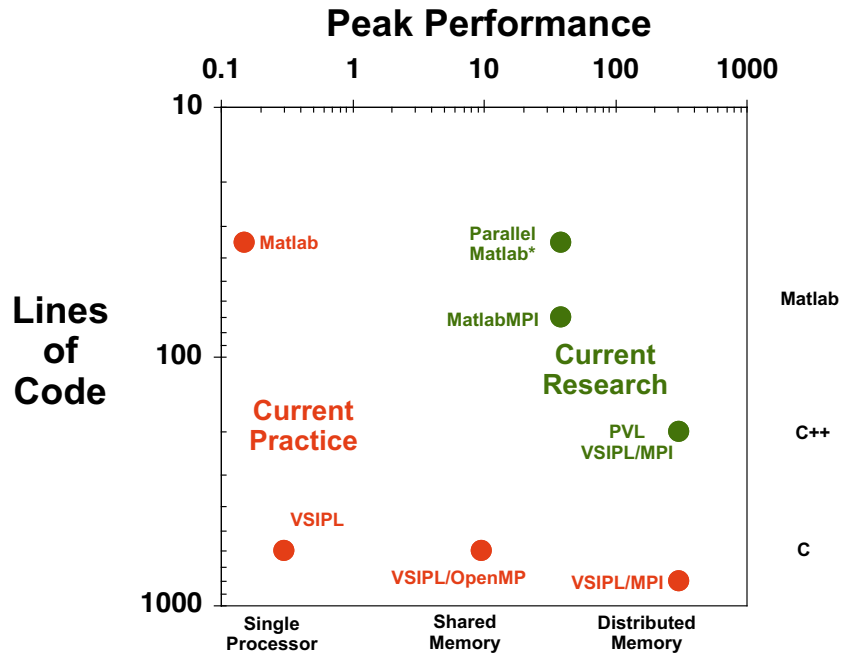}}
\caption{ {\bf Productivity vs Performance.}
  Lines of code as a function maximum achieved performance (measured
in units of single processor theoretical peak) for different
implementations of the same image filtering application.  Higher level
languages allow the same application to be implemented with fewer
lines of code.  Increasing the maximum performance generally increases
the lines of code.  MatlabMPI allows high level languages to run on
multiple processors.  Parallel Matlab is an estimate of what will be
possible using the Parallel Matlab Toolbox.  Abbreviations: VSIPL
(Vector, Signal, and Image Processing Library) see www.vsipl.org; PVL
(Parallel Vector Library) see www.hpec-si.org.
}
\label{fig:Prod_vs_Perf}
\end{figure}

  The ultimate goal of running Matlab on parallel computers is to
increase programmer productivity and decrease the large software cost of
using HPC systems.
Figure~\ref{fig:Prod_vs_Perf}) plots the software cost
(measured in Software Lines of Code or SLOCs) as a function of the
maximum achieved performance (measured in units of single processor
peak) for the same image filtering application implemented using
several different libraries and languages (VSIPL, MPI, OpenMP, using
C++, C, and Matlab \cite{Kepner2002}).  These data show
that higher level languages require fewer lines to implement
the same level of functionality.  Obtaining increased peak
performance (i.e. exploiting more parallelism) requires more lines of
code.  MatlabMPI is unique in that it achieves a high peak performance
using a small number of lines of code.

  Two useful metrics we have developed for measuring software
productivity are Gigaflops/SLOC and CPUs/SLOC.  Gigaflops/SLOC will
increase in a given application if it can be made to run faster or can
be implemented with fewer lines of code.  CPUs/SLOC will increase if
the number of processors that can be effectively used in an
application can be increased without increasing the number of lines of
code.  The test application does extremely well in both of these
measures, achieving 0.85 Gigaflops/SLOC and 4.4 CPUs/SLOC, both of
which are the highest values recorded for any application.

\section{Conclusions and Future Work}

  The use of file I/O as a parallel communication mechanism is not new
and is now increasingly feasible with the availability of low cost high
speed disks.  The extreme example of this approach are the now popular
Storage Area Networks (SAN), which combine high speed routers and disks
to provide server solutions. Although using file I/O increases the
latency of messages it normally will not effect the bandwidth. 
Furthermore, the use of file I/O has several additional functional
advantages which make it easy to implement large buffer sizes,
recordable messages, multi-casting, and  one-sided messaging.  Finally,
the MatlabMPI approach is readily applied to any language  (e.g. IDL,
Python, and Perl).

   MatlabMPI demonstrates that the standard approach to writing
parallel programs in C and Fortran (i.e. using MPI) is also valid in
Matlab.  In addition, by using Matlab file I/O, it was possible to
implement MatlabMPI entirely within the Matlab environment, making it
instantly portable to all computers that Matlab runs on.  Most
potential parallel Matlab applications are trivially parallel and
don't require high performance.  Never-the-less, MatlabMPI can match C
MPI performance on large messages.  The simplicity and performance of
MatlabMPI makes it a very reasonable choice for programmers that want
to speed up their Matlab code on a parallel computer.

\begin{figure}[tbh]
\centerline{\includegraphics[width=4.5in]{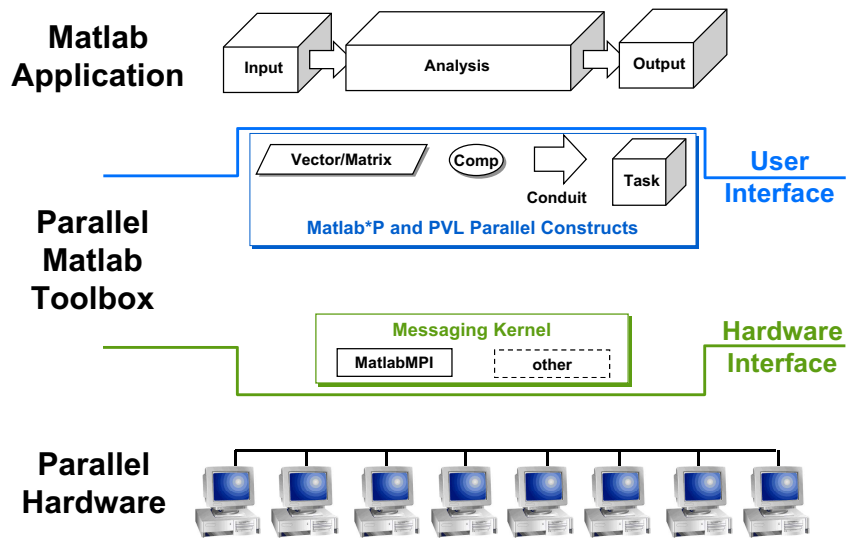}}
\caption{ {\bf Future Layered Architecture.}
  Design of the Parallel Matlab Toolbox which will create
distributed data structures and dataflow objects built on
top of MatlabMPI (or any other messaging system).  
}
\label{fig:LayeredArch}
\end{figure}

  MatlabMPI provides the highest productivity parallel computing
environment available.  However, because it is a point-to-point
messaging library, a significant amount code of must be added to any
application in order to do basic parallel operations.  In the test
application presented here, the number of lines of Matlab code
increased from 35 to 70.  While a 70 line parallel program is
extremely small, it represents a significant increase over the single
processor case.  Our future work will aim at creating higher level
objects (e.g. distributed matrices) that will eliminate this parallel
coding overhead (see Figure~\ref{fig:LayeredArch}).  The resulting
``Parallel Matlab Toolbox'' will be built on top of the MatlabMPI
communication layer, and will allow a user to achieve good parallel
performance without increasing the number lines of code.

%
%

\section*{Acknowledgments}

The authors would like to thank the sponsorship of the DoD High Performance
Computing Modernization Office, and the staff at the supercomputing
centers at Boston University, Ohio State University and the Maui High
Performance Computing Center.

\appendix

\section{Parallel Image Filtering Test Application}

\begin{verbatim}
MPI_Init;  % Initialize MPI.
comm = MPI_COMM_WORLD;  % Create communicator.
comm_size = MPI_Comm_size(comm);  % Get size.
my_rank = MPI_Comm_rank(comm);  % Get rank.
% Do a synchronized start.
starter_rank = 0;
delay = 30;  % Seconds
synch_start(comm,starter_rank,delay);
n_image_x = 2.^(10+1)*comm_size;  % Set image size (use powers of 2).
n_image_y = 2.^10;
n_point = 100;  % Number of points to put in each sub-image.
% Set filter size (use powers of 2).
n_filter_x = 2.^5;
n_filter_y = 2.^5;
n_trial = 2;  % Set the number of times to filter.
% Computer number of operations.
total_ops = 2.*n_trial*n_filter_x*n_filter_y*n_image_x*n_image_y;
if(rem(n_image_x,comm_size) ~= 0)
 disp('ERROR: processors need to evenly divide image');
 exit;
end
disp(['my_rank: ',num2str(my_rank)]);  % Print rank.
left = my_rank - 1;  % Set who is source and who is destination.
if (left < 0)
  left = comm_size - 1;
end
right = my_rank + 1;
if (right >= comm_size)
  right = 0;
end
tag = 1;  % Create a unique tag id for this message.
start_time = zeros(n_trial);  % Create timing matrices.
end_time = start_time;
zero_clock = clock;  % Get a zero clock.
n_sub_image_x = n_image_x./comm_size;  % Compute sub_images for each processor.
n_sub_image_y = n_image_y;
% Create starting image and working images..
sub_image0 = rand(n_sub_image_x,n_sub_image_y).^10;
sub_image = sub_image0;
work_image = zeros(n_sub_image_x+n_filter_x,n_sub_image_y+n_filter_y);
% Create kernel.
x_shape = sin(pi.*(0:(n_filter_x-1))./(n_filter_x-1)).^2;
y_shape = sin(pi.*(0:(n_filter_y-1))./(n_filter_y-1)).^2;
kernel = x_shape.' * y_shape;
% Create box indices.
lboxw = [1,n_filter_x/2,1,n_sub_image_y];
cboxw = [n_filter_x/2+1,n_filter_x/2+n_sub_image_x,1,n_sub_image_y];
rboxw = [n_filter_x/2+n_sub_image_x+1,n_sub_image_x+n_filter_x,1,n_sub_image_y];
lboxi = [1,n_filter_x/2,1,n_sub_image_y];
rboxi = [n_sub_image_x-n_filter_x/2+1,n_sub_image_x,1,n_sub_image_y];
start_time = etime(clock,zero_clock);  % Set start time.
% Loop over each trial.
for i_trial = 1:n_trial
  % Copy center sub_image into work_image.
  work_image(cboxw(1):cboxw(2),cboxw(3):cboxw(4)) = sub_image;
  if (comm_size > 1)
    ltag = 2.*i_trial;    % Create message tag.
    rtag = 2.*i_trial+1;
    % Send left sub-image.
    l_sub_image = sub_image(lboxi(1):lboxi(2),lboxi(3):lboxi(4));
    MPI_Send(  left, ltag, comm, l_sub_image );
    % Receive right padding.
    r_pad = MPI_Recv( right, ltag, comm );
    work_image(rboxw(1):rboxw(2),rboxw(3):rboxw(4)) = r_pad;
    % Send right sub-image.
    r_sub_image = sub_image(rboxi(1):rboxi(2),rboxi(3):rboxi(4));
    MPI_Send( right, rtag, comm, r_sub_image );
    % Receive left padding.
    l_pad = MPI_Recv( left, rtag, comm );
    work_image(lboxw(1):lboxw(2),lboxw(3):lboxw(4)) = l_pad;
  end
  work_image = conv2(work_image,kernel,'same');   % Compute convolution.
  % Extract sub_image.
  sub_image = work_image(cboxw(1):cboxw(2),cboxw(3):cboxw(4));
end
end_time = etime(clock,zero_clock);  % Get end time for the this message.
total_time = end_time - start_time  % Print the results.
% Print compute performance.
gigaflops = total_ops / total_time / 1.e9;
disp(['GigaFlops: ',num2str(gigaflops)]);
MPI_Finalize;  % Finalize Matlab MPI.
exit;
\end{verbatim}


\newpage


\end{document}